\documentstyle[12pt,blois,psfig]{article}
\begin{document}
\heading{EXTREMELY RED GALAXIES: OLD OR DUSTY ?} 

\author{A. Cimatti $^{1}$, P. Andreani $^{2}$, H.J.A. R\"ottgering $^{3}$, R.
Tilanus $^{4}$} 
{$^{1}$ Osservatorio Astrofisico di Arcetri, Firenze, Italy} 
{$^{2}$ Dipartimento di Astronomia, Padova, Italy \& MPE, Garching, Germany} 
{$^{3}$ Sterrewacht, Leiden, The Netherlands} 
{$^{4}$ JACH, Hawaii, USA \& NFRA, Dwingeloo, The Netherlands}

\begin{bloisabstract}

The nature of the extremely red galaxies is poorly known, the critical 
issue being whether they are old ellipticals at $z>1.5$ or distant star-forming 
galaxies strongly reddened by dust extinction. Here we review the little 
we know about them, and we present recent and preliminary results from 
our ongoing submillimeter -- millimeter observations.

\end{bloisabstract}

\section{Introduction}

Recent deep optical observations discovered a population 
of high-$z$ star-forming galaxies \cite{steidel,hu98} and, together with the 
spectroscopic surveys of galaxies at lower $z$, these results have been 
used to trace back the history of the global star formation of the Universe 
\cite{madau}. However, these galaxies have been selected at optical wavelengths 
according to criteria (such as the presence of a Ly-break continuum and of the 
Ly$\alpha$ emission) that are vulnerable to selection effects like those caused by 
dust extinction. Therefore, one of the main questions is whether the optical 
surveys have not missed significant populations of distant galaxies. If so, 
this could seriously change our understanding of the global properties of 
the whole population of high-$z$ galaxies. 

An example of a class of objects that are missed by traditional optical selection 
techniques to search for distant objects are extremely red galaxies (ERGs), 
enigmatic faint objects that appear in combined optical and infrared surveys  
\cite{hu94,mc92}. These 
galaxies, sometimes even undetected in deep optical images, are found both 
in random sky fields and around high-$z$ AGN. Their sky surface density
in the field at $K<20$, although still poorly known, is of the order of 
$\approx$0.1-0.2 arcmin$^{-2}$ for $R-K>6$ or $I-K>4$, and $\approx$0.01-0.02 
arcmin$^{-2}$ for $R-K>7$ or $I-K>5$ \cite{barg99,hu94,co94,co96,mcleod95,
thom98}. It should be noted that, in case of non-resolved objects, some 
contamination by dM stars and brown dwarfs can be present (see \cite{thom98}). 
ERGs are too faint to obtain redshifts with 4-m class 
telescopes, and only the Keck 10m telescope provided 
so far the only redshift available for one of these galaxies (HR10, $z=1.44$) 
\cite{gd96}. Although their colors and faintness indicate 
that these galaxies are at $z>1$, their nature is ambiguous because their 
colors are consistent with two very different scenarios : {\sl (1)} they 
are ``old'' galaxies (age $>$ 1 Gyr) at $z>1.5$, the colors being so red 
because of lack of star formation and because of the strong K-correction 
effect. In this case, ERGs would represent the oldest envelope of the
distant galaxy population, and they could provide crucial information
on the earliest epoch of galaxy formation and on the evolution of the
elliptical galaxies; {\sl (2)} they are dusty and star-forming galaxies 
where the extreme colors are the result of a severe reddening of the 
rest-frame UV continuum. In this case, ERGs are important to constrain the 
space density of dusty star-forming galaxies in the early Universe. It is 
evident that both scenarios are very relevant to galaxy formation 
and evolution. 

\section{Our projects}

We started a multi-wavelength study aimed at unveiling the nature of ERGs
through optical to millimeter observations. Two small surveys have been 
carried 
out in order to select two complete samples of ERGs: one in random sky fields 
and one around radio-loud AGN at $z>1.5$. The aims of these surveys are to 
constrain the abundance of ERGs and to provide targets for VLT spectroscopy. 
These surveys made use of ESO and HST imaging data. Preliminary results of 
one of the surveys are described by Pozzetti et al. (these proceedings).

Another project makes use of submillimeter and millimeter continuum
observations in order to test the hypothesis that ERGs are dusty 
starburts, i.e.  galaxies where the UV continuum from OB stars is 
heavily absorbed by dust and re-emitted in the rest-frame far-IR. The 
observations are performed with the JCMT 15m telescope equipped with 
SCUBA \cite{scuba} working at 450$\mu$m and 850$\mu$m, and with the 
IRAM 30m telescope 
equipped with the MPIfR 1.25mm bolometer array. The main purposes of this 
project are to search for the redshifted dust thermal emission, to measure the 
rest-frame far-IR luminosities, to estimate the star formation rates (SFRs), 
and to investigate if ERGs are related to the population of dusty
submm-selected galaxies recently discovered in deep observations of
``blank'' fields by \cite{ivison,hughes,barger}. The ultimate aim is to 
investigate whether they contribute significantly to 
the global star formation history of the Universe
and to the recently discovered cosmic far-IR -- mm background.

\section{Submillimeter -- millimeter results}

Our observations allowed us to detect for the first time submm-mm
continuum emission from an ERG (HR10) \cite{cima98}. HR10 ($z=1.44$, 
\cite{hu94, gd96,cima97}) was detected independently with the JCMT+SCUBA 
($\lambda_{obs}=850\mu$m) and the IRAM 30m telescope ($\lambda_{obs}=
1250\mu$m) (Fig. 1). 

The most plausible mechanism to explain the detected submm-mm continuum 
is thermal dust emission. For dust temperatures $T_{d}\sim$30-45 K, 
the total dust mass is $M_{dust}\approx 8-4\times 10^{8} \ 
h_{50}^{-2}$ M$_{\odot}$ ($q_0=0.5$, $\beta=2$), and the rest-frame
far-IR luminosity is $L_{FIR} \approx 2-2.5\times 10^{12} \ h_{50}^{-2}$
L$_{\odot}$ ($\Delta\lambda_{rest}=10-2000\mu$m, $q_0=0.5$). This 
luminosity places HR10 in the class of ultra-luminous infrared
galaxies (ULIGs, \cite{sanders}). Adopting the relationship 
$SFR = (0.8 \div 2.1) 10^{-10}(L_{FIR}/$L$ _{\odot})$ [M$_{\odot}$yr$^{-1}$], 
and assuming no AGN contribution, the resulting star formation 
rate of HR10 is in the range of $200-500 \ h_{50}^{-2}$ M$_{\odot}$yr$^{-1}$. 
It is striking to notice that the SFRs estimated from the H$\alpha$ luminosity
($\sim 80 \ h_{50}^{-2}$ M$_{\odot}$ yr$^{-1}$) and from the UV continuum 
luminosity ($\sim 1 \ h_{50}^{-2}$ M$_{\odot}$yr$^{-1}$) are underestimated by 
a factor up to $\approx$500 because of the strong dust extinction that
obscures the rest-frame UV-optical light.

A sample of other 8 ERGs (excluding HR10) with $K<20$ and $I-K>6$ has been 
observed so far. The final data reduction is currently under way. 
For 4 ERGs we could reach the sensitivity required by our survey (rms$<$2 
mJy at 850$\mu$m), whereas the weather was not good enough to obtain deep
data at 450$\mu$m. A preliminary analysis suggests that we obtained two 
marginal detections at 850$\mu$m that need to be confirmed with deeper
observations. We have also searched for the presence of a positive signal 
from the population of ERGs by coadding the data of the whole sample. The
weighted average of the 850$\mu$m flux density of the entire sample 
provides a signal at $3\sigma$ level. The significance increases to 
$4\sigma$ level if only the 6 objects with a positive signal are selected. 
Together with the detection of HR10, this clearly hints that the colors of 
a large fraction of the ERGs are severely affected by dust. We recall 
that these findings still need to be confirmed and that the final results 
will be presented in a forthcoming paper (Cimatti et al., in preparation).

It should be recalled here that the physical interpretation of the
submm-mm observations is hampered by the fact that the redshifts
of the ERGs are unknown (with the exception of HR10). However, thanks
to the strong K-correction effect caused by the steep grey-body dust 
spectra, the expected flux at $\lambda_{obs}=850\mu$m of a dusty 
star-forming galaxy at $1<z<10$ is not a strong function of the redshift. 
Thus, since ERGs are expected to be at $z>1$, a detection at a few mJy level
at $\lambda_{obs} =850\mu$m directly implies a large content of dust and 
high $L_{FIR}$ and 
SFR irrespective of the redshift (see also \cite{hughes,barger}). 

\begin{figure}
\centerline{
\psfig{figure=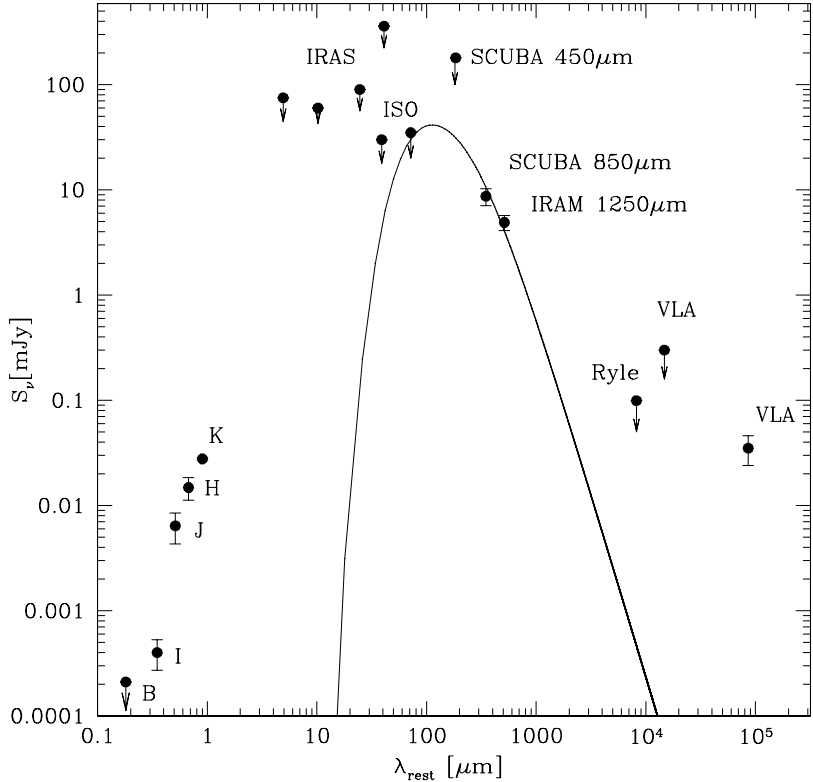,width=12cm}
}
\caption{\label{fig1}
The optical-to-radio spectral energy distribution of HR10 \cite{cima98} 
and a grey-body curve ($T_{d}$=45 K) consistent with the SCUBA and IRAM 
detections and with the ISO upper limit.
}
\end{figure}

\section{Implications and future}

The main implications of our results can be summarized as follows:
{\sl (1)} the submm-mm observations {\it prove} that at least some of
the ERGs are infrared (ultra)luminous dusty galaxies undergoing a major 
burst of star formation. {\sl (2)} our results also clearly show that 
the optical observations only do not sample 
the whole population of high-$z$ galaxies: a galaxy
such as HR10 would not be selected in the optical as a Lyman-break
galaxy because of its extremely red spectrum; {\sl (3)} the combination of
deep optical and near-IR imaging coupled with submm photometry
provides a successful technique to select and study galaxies 
that, despite their virtually identical colors, represent two 
opposite stages of galaxy evolution: old passively evolved 
systems and dusty star-forming galaxies.

Three main questions should be answered in order to characterize
the nature of the ERGs: {\sl (1)} what is the fraction of
dusty starbursts ? {\sl (2)} what is their space density 
in the Universe ? {\sl (3)} what are the ages of the putative
old ellipticals that populate the other fraction of ERGs ?

These wide open problems can be addressed with deep optical
and near-IR spectroscopy with 8-10m class telescopes aimed
at deriving the redshifts and the spectral types of the ERGs.
High-quality spectra will provide clues on the ages of the 
stellar population(s) in the case of old ellipticals, and they 
will allow us to better constrain the earliest epoch of 
galaxy formation and the permitted values of $H_0$ and $q_0$. 
Finally, the knowledge of the redshift distribution and of the 
volume density of the star-forming ERGs would allow us to understand 
their contribution to the global star-formation history
of the Universe and to the cosmic far-IR -- mm background.

\acknowledgements{The JCMT is operated by the Joint Astronomy
Centre on behalf of the Particle Physics and Astronomy Research Council
of the United Kingdom, the Netherlands Organisation for Scientific
Research, and the National Research Council of Canada.
This work was supported in part by the Formation and Evolution of
Galaxies network set up by the European Commission under contract ERB
FMRX-- CT96--086 of its TMR programme.
HJAR acknowledges support from a programme subsidy granted by the
Netherlands Organisation for Scientific Research (NWO).}


%
\vfill
\end{document}